\documentstyle[prl,aps,twocolumn]{revtex}
\begin{document}
\draft
\flushbottom

\title{\bf Atomic Layering at the Liquid Silicon Surface:
a First-Principles Simulation}

\author{
Gabriel Fabricius$^{1,2}$,
Emilio Artacho$^{1,3}$,
Daniel S\'anchez-Portal$^1$,
Pablo Ordej\'on$^4$,
D. A. Drabold$^5$,
and Jos\'e M. Soler$^1$
}

\address{
$^1$Dpto. de F\'{\i}sica de la Materia Condensada
    and Instituto Nicol\'as Cabrera,
    Univ. Aut\'onoma, E-28049 Madrid, Spain \\
$^2$Dpto. de F\'{\i}sica, Univ. Nacional de La Plata,
    1900 La Plata, Argentina \\
$^3$P\^ole Scientifique de Mod\'elisation Num\'erique,
   Ecole Normale Sup\'erieure de Lyon, 69364 Lyon, Cedex 07, France \\
$^4$Dpto. de F{\'{\i}}sica, Universidad de Oviedo, 33007 Oviedo, Spain
\\
$^5$Dep. of Physics and Astronomy, Ohio University,
    Athens, OH 45701-2979
}

\date{\today}
\maketitle

\begin{abstract}
We simulate the liquid silicon surface with first-principles
molecular dynamics in a slab geometry.
We find that the atom-density profile
presents a pronounced layering, similar to those observed in
low-temperature liquid metals like Ga and Hg.
The depth-dependent pair correlation function shows that the effect
originates from directional bonding of Si atoms at the surface,
and propagates into the bulk.
The layering has no major effects in the electronic and
dynamical properties of the system, that are very similar to those of
bulk liquid Si.
To our knowledge, this is the first study of a liquid surface by
first-principles molecular dynamics.
\end{abstract}


\narrowtext
Liquid metal surfaces have attracted much attention during the last
years
\cite{nature,science,Hg_exp,Ga_exp,rice98,tosatti}.
Their particular properties, very different from those of non-metals,
have been investigated by {\AA}ngstr\"om-resolution experiments,
and simulated by different approaches.
One of their most interesting features is the atomic layering,
a density oscillation
that originates at the sharp liquid-vapor interface,
and extends several atomic diameters into the bulk.
Although some experiments supported an increased
surface density at $l$-Hg \cite{nature}, and some kind
of atomic layering was predicted theoretically \cite{1sttheor},
its existence has been demonstrated unambiguously
only recently by X-Ray reflectivity in Hg \cite{Hg_exp},
Ga \cite{Ga_exp,Ga_expT} and Ga-In alloys \cite{GaIn_exp}.
Most of the experiments have been done close to the low melting
temperatures of these metals,
but Regan {\it et al} \cite{Ga_expT} studied the Ga surface
up to 170$^{\circ}$C.
They found that capilarity waves strongly decrease the reflectivity
peak heights, but not the peak widths, suggesting that the
decay length of local layering is temperature-independent, and that
surface layering is still present quite above the melting point.

Different explanations have been proposed to account for this effect
\cite{Lai}.
Rice {\it et al} \cite{rice98} have argued that the abrupt decay of
the delocalized electron density forms a flat potential barrier against
which the ions lay orderly, like hard spheres against a hard wall.
Tosatti {\it et al} \cite{tosatti} have used the glue model
of metallic cohesion to argue that surface atoms,
trying to effectively recover their optimal coordination,
alternatively increase and decrease their density.
Surface layering effects, like surface-enhanced smectic ordering,
have also been observed in liquid crystals \cite{molecularLiquids}.
In this case, its origin is the tendency of the highly nonspherical
molecules to present a particular orientation towards the surface.

Liquid silicon ($l$-Si) is a rather peculiar system.
Silicon transforms, at 1684 K, from a covalent semiconductor solid,
with diamond structure and coordination 4, to a liquid metal.
Experiments \cite{Si_exp1,Si_exp2} and MD simulations \cite{carr,cheli}
show that its coordination ($\sim 6-7$) is lower
than that of typical liquids ($\sim 12$), due to
the persistence of directional bonding in the liquid phase.
In spite of the enormous literature on its solid surfaces,
very little is known on the structure of the liquid silicon surface.
Measurements are very difficult because
of its high reactivity and melting temperature.
Model calculations, and computer simulations
with semiempirical potentials, are also difficult because of the
mentioned coexistence of covalent and metallic bonding,
and its unknown interplay at the surface.

In this letter we present a study of the $l$-Si surface by
first-principles molecular dynamics (MD) simulation
\cite{CarParrinello}.
This approach deals equally well with covalent and metallic bonding,
and it is therefore very well suited for this problem.
Electrons are treated by solving the Kohn-Sham \cite{KohnSham}
equations selfconsistently for each ionic configuration,
using the local density approximation for exchange and correlation.
The quantum mechanically obtained forces are then used to generate
the classical trajectories of the ion cores.
The calculations were performed with the SIESTA program \cite{SIESTA}
using a linear combination of numerical atomic orbitals as the basis
set,
and norm-conserving pseudopotentials \cite{TroullierMartins}.
A uniform mesh with a planewave cutoff of 40 Ry is used to represent
the electron density, the local part of the pseudopotential,
and the Hartree and exchange-correlation potentials.
Only the $\Gamma$ $k$-point was used in the simulations,
since previous work \cite{Sispin} found cell-size effects to be small.
For the present calculation we used a minimal basis set of four orbitals
(1 $s$ and 3 $p$) for each Si atom, with a cutoff radius of 2.65 \AA.
We have extensively checked the basis with static calculations
of different crystalline Si phases and solid surfaces, and MD
simulations
of the bulk liquid \cite{longwork}.
The energy differences between solid phases are described within 0.1 eV
of other {\it ab-initio} calculations.
The diamond structure has the lowest energy, with a lattice
parameter of 5.46 \AA~ (0.5 $\%$ larger than the experimental value).
Adatom- and dimer-based (111) and (100) surface reconstructions 
found in other {\it ab-initio} calculations \cite{Vanderbilt-Northrup}
are well reproduced, with geometries and relative energies changing
less than $\sim 0.1$ \AA~ and $\sim$0.15 eV when moving from a 
$\Gamma$-point calculation with a minimal basis set, to a converged 
$k$-sampling with double-$\zeta$ and polarization orbitals.
The structural, electronic and dynamical properties of $l$-Si
are in good agreement with other {\it ab-initio} calculations
at the same density and temperature \cite{carr,cheli}.
The calculated diffusion constant ($1.5 \times 10^{-4}$ cm$^2$/s)
is somewhat smaller than that obtained with a double-$\zeta$ or
polarized basis ($1.7-2.0 \times 10^{-4}$ cm$^2$/s)
which is in agreement with other {\it ab-initio} simulations.
We interpret that the minimal basis overestimates the energies of
the saddle point configurations occurring during diffusion, but we
consider that this is not critical for the present application.
Also, we leave for a future work the inclusion of spin fluctuations,
which affect significantly the diffusion constant but not the
structural properties of the liquid \cite{Sispin}.

We first perform a long simulation of bulk $l$-Si at $T$=1800 K
\cite{constT},
using a cubic 64-atom cell with periodic boundary conditions.
The fixed cell size (10.58 \AA) was adjusted to obtain zero mean
pressure, and corresponds to a density 3$\%$ smaller than the
experimental density near the melting point.
We then construct our initial 96-atom slab by repeating
one bulk unit cell in the $x$ and $y$ directions, and one and a half
cells in the $z$ direction, plus 10 \AA~ of vacuum.
No particles leave the slab during the 30 ps simulation.
After a relaxation of 10 ps, the system reaches equilibrium and the
averaged magnitudes are essentially the same for
the next or the last 10 ps, and for both sides of the slab.
These long relaxation and observation times are required
because the calculated density autocorrelation time at the surface
($\sim 1$ ps) is considerably longer than the typical
bulk-liquid correlation times ($\sim$ 0.1 ps) \cite{carr}.
The average surface energy (836 $\pm$ 40 dyn/cm) is in good agreement
with the experimental surface tension (850 dyn/cm at 1800 K) \cite{iida},
suggesting a small entropic contribution.

In Fig.~\ref{fig1} (solid line) we present the ionic density profile
$\rho(z)$.
It shows a pronounced atomic layering, with similar features as
those reported for the Ga surface \cite{Ga_exp}.
Like in that case, $\rho(z)$ can be fitted accurately by
a sharp error function at the surface, and a sinusoidal wave with
an exponential decay towards the bulk:
by superimposing two such functions (not shown), for both surfaces,
we obtain similar values of the parameters and
an oscillation period of 2.5 \AA.
To check that the observed layering is not a sign of incipient
crystallization, we have computed several magnitudes
in the central region of the slab ($|z| < 3.7$ \AA).
The radial and angular distribution functions,
electronic and vibrational densities of states,
and the diffusion constant, are all very similar to those
of bulk $l$-Si, and have no resemblance of those in the solid phases.
As an example, we compare in Fig.~\ref{fig2} the bond-angle
distribution function for the bulk and slab simulations,
using a bond cutoff distance of $r_m$=3.10 \AA \cite{rmin}.
For a better understanding of the origin of the layering,
we compute the normalized density-density correlation function
\cite{TildesleyAllen}:
\begin{equation}
c_{\rho}(z_0,z)=
   \frac{ \langle \delta\rho(z_0,t) \delta\rho(z,t) \rangle }
        { \langle \delta\rho^2(z_0,t) \rangle^{\frac{1}{2}}
          \langle \delta\rho^2(z,t)   \rangle^{\frac{1}{2}} },
\end{equation}
where $\langle \rangle$ denotes time average and $\delta\rho$
is the difference between the instantaneous density at time $t$,
$\rho({\bf r},t) = \sum_{i=1}^N \delta({\bf r}-{\bf r}_i(t))$,
and the average density $\rho({\bf r})=\langle \rho({\bf r},t) \rangle$,
where $\delta({\bf r})$ is Dirac's function.
Fig.~\ref{fig1} also shows $c_{\rho}(z_0,z)$ for $z_0$ at the
positions of the outermost peaks of each side.
Its decaying oscillation is clearer than that
of the density profile, all whose relevant features match
very well with the superposition of the two $c_{\rho}$'s.
The apparent lack of decay of $\rho(z)$ towards the
interior is peculiar to our particular slab thickness,
because the superposition is positive at the center of the slab,
and negative at the two surfaces.
Most important is, however, that the two surface-induced oscillations
are clearly independent of each other ($c_{\rho}$'s out of phase),
and incommensurate to the slab thickness.
The density layering is thus an intrinsic
surface feature and not a result of finite size effects.

In order to obtain information about the bond orientations at the
surface
we calculate the two-particle density:
\begin{equation}
\label{rhor0r}
\rho_2({\bf r}_0;{\bf r}) = \frac{N/(N-1)}{\rho({\bf r}_0)}
     \left\langle \sum_{i=1}^{N} \sum_{j \neq i}
     \delta({\bf r}_0-{\bf r}_i) \delta({\bf r}-{\bf r}_j)
\right\rangle.
\end{equation}
To represent $\rho_2({\bf r}_0;{\bf r})$, we first average over the
directions parallel to the surface:
\begin{eqnarray}
\label{rhoz0zx}
\rho_2(z_0;z,x) &&
 = \frac{1}{2 \pi x A} \int \int d^3{\bf r}'_0 d^3{\bf r}'
     \rho_2({\bf r}'_0;{\bf r}') \delta(z_0-z'_0) \nonumber \\
&& \delta(z-z') \delta(x-\sqrt{(x'-x'_0)^2+(y'-y'_0)^2}),
\end{eqnarray}
where $A$ is the area of the simulation cell.
In Fig.~\ref{fig3} we show $\rho_2(z_0;z,x)$ for
$z_0$ located at the three peaks of $\rho(z)$.
Fig. \ref{fig3}(a) shows a clear tendency of surface atoms
to form bonds parallel and normal to the surface.
The height of the correlation peaks goes well beyond
those of the density, which can be seen also in the figure
(notice that $\rho_2({\bf r}_0;{\bf r}) \rightarrow \rho({\bf r})$ for
$|{\bf r - r}_0| \rightarrow \infty$).
This shows that the bond-induced correlations are responsible for
the layering of the density, and not the other way around.
Fig. \ref{fig3}(b) shows a similar, but attenuated tendency,
that disappears in the third layer (Fig.~\ref{fig3}(c)),
which already has a very symmetric, bulk-like pair correlation function.

Further insight can be obtained from the $z$-dependent coordination
$n(z)$, defined as the average number of neighbors
within a distance $r_m$.
At the bulk, we obtain $n$=6.4, in
agreement with the experimental value \cite{Si_exp1}.
The distribution of local coordinations (DLC) is also very close
to those of other {\it ab-initio} calculations \cite{carr,cheli},
showing a maximum at coordination 6.
We can also use the bulk simulation to construct an ideally terminated
surface, cutting abruptly the system at, say, $z$=0.
We then find $n(z)=4.3$ at $z=1.0$ \AA, which is the distance between
the
outermost peak and the inflection point in the slab density profile.
At the outermost peaks of the actual slab,
we obtain $n(z)$=5.3, and a DLC peaked at 5.
These values show that surface structural rearrangements
increase the coordination of the ideally terminated surface,
reaching a value of only one neighbor less than in the bulk.
If we associate coordination 6, in the bulk liquid, with an octahedral
arrangement, a simple picture can be drawn, in which the surface atoms
try to preserve their bulk environment while minimizing the number of
broken bonds.
As a consequence, the octahedra get oriented in the surface so that
only one broken `bond' points towards vacuum, with another bond towards
the interior and four bonds laying on the surface plane.
This picture is consistent with figures \ref{fig3}(a) and \ref{fig3}(d).
In the latter, we have restricted the sum over $i$ in
eq.~(\ref{rhor0r}) to particles having coordination 5,
what results in even more pronounced peaks
in the $x$ and $z$ directions.
Also, we note that the maximum of $\rho_2(z_0;z,0)$ occurs at
$|z-z_0|=2.5$ \AA,
what explains the same period observed in the density profile.
The same distance is found for the in-plane surface bonds
(i.e.~for $\rho_2(z_0;z_0,x)$), and for the bulk bonds.
Thus, contrary to other metals, we do not find a shortening
of the surface bonds, and the silicon surface layering seems to be
related only to the bond orientations.
However, it must be emphasized that the bond angle distribution
(Fig.~\ref{fig2}) is very wide, indicating a large variety of
fluctuating
atomic environments \cite{carr}, so that our `oriented octahedra' should
be considered only as a very rough and qualitative picture.

 An interesting question is whether the surface structural
rearrangements
 produce a noticeable signature in the electronic structure.
 In Fig.~\ref{fig4}, we compare the local density of states (LDOS)
 at the outermost peaks of $\rho(z)$ and at the center of the slab.
 Although $k$-sampling is important for a converged LDOS
 \cite{longwork,Hafner}, we use here only the $\Gamma$-point
 eigenvalues, to facilitate the comparison with previous work
 \cite{carr,cheli} and because we focus on its spatial variation.
 It can be seen that, apart from a slight narrowing, due to the
 diminished surface coordination, there are no major differences,
 suggesting that the surface and bulk atomic environments are
 rather similar, and again pointing towards bond
 orientations as responsible for surface layering.

In conclusion, we have performed the first study of a liquid
surface by first-principles MD simulation.
In spite of the high melting temperature of Si,
we find a marked layering of the density near the surface,
similar to those observed in other metals, like Ga and Hg,
with low melting temperatures.
However, the surface layering of Si seems to have an origin at
least partially different from that in other metals, with
remanent directional covalent bonding playing an essential role.
In spite of the rather slow decay of the layering towards the
bulk, the average structural, dynamical, and electronic properties
converge very rapidly to their bulk liquid values.
Although more converged simulations would be highly desirable in 
the future, we consider that this work provides a new qualitative 
understanding of the complex structure of liquid surfaces.

We acknowledge useful discussions with E. Chac\'on and M. Weissmann.
This work was supported by
Argentina's CONICET and by Spain's DGES grant PB-0202.

\begin{figure}
\caption{Solid line: density profile $\rho(z)$
(relative to the bulk density $\rho_0$) of a liquid-Si slab,
averaged during 20 ps and smeared by a gaussian convolution of 0.15 \AA.
Dashed ($z_0$=6.2 \AA) and dotted ($z_0$=-6.2 \AA) lines:
density-density correlation function $c_{\rho}(z_0,z)$ (eq.~(1)).
We use a centered window of 0.5 \AA~ for $z_0$,
and a gaussian smearing of 0.5 \AA~ for $z-z_0$.}
\label{fig1}
\end{figure}

\begin{figure}
\caption{Distribution of bond-angles in the central region of
the slab (continuos line) and in bulk $l$-Si (dashed line).}
\label{fig2}
\end{figure}

\begin{figure}
\caption{Two-particle density $\rho_2(z_0;z,x)$ (eq.~(3)) for:
(a) $z_0=6.2$ \AA; (b) $z_0=4.0$ \AA; and (c) $z_0=1.3$ \AA.
The `volcano' is centered at $x=0, z=z_0$ 
(position of the reference particle). 
$\rho_2$ has been extended symmetrically to $x<0$ to facilitate 
its visualization (a line of ripples is produced by noise 
due to poorer statistics in eq.~(3) at $x \simeq 0$).
(d) The same as (a), but restricted to atoms at $z_0$
having coordination 5.}
\label{fig3}
\end{figure}

\begin{figure}
\caption{Local density of electron states of the atoms in the
outermost density peaks (solid line),
and of those at the center of the slab (dashed line).}
\label{fig4}
\end{figure}


\begin{references}
\bibitem{rmin}
   We use for $r_m$ the position of the minimum of the pair
   correlation function $r^2 g(r)$ in bulk $l$-Si.
\end{references}
\end{document}